\documentclass[aps,twocolumn,prx,preprintnumbers,showpacs,amsmath,amssymb,superscriptaddress]{revtex4-1}
\usepackage[pdftex, colorlinks, citecolor=blue]{hyperref}   
\usepackage{graphicx}
\usepackage{dcolumn}
\usepackage{multirow}
\usepackage{booktabs}
\usepackage{txfonts}
\usepackage{xcolor}
\usepackage{bm}
\usepackage{amssymb}
\usepackage{amsmath}
\usepackage{latexsym}
\usepackage{epsfig}
\usepackage{amsbsy}
\usepackage{array}
\usepackage{tabularx}
\usepackage{tabularray}
\usepackage{threeparttable}
\usepackage{esvect} 
\usepackage{extarrows}
\usepackage{float}
\usepackage[british]{babel}

\begin{document}
\title{Finite-temperature ductility-brittleness and electronic structures of Al$_{n}$Sc (n=1, 2 and 3)}

\author{Xue-Qian Wang}
\affiliation{%
Key Laboratory for Anisotropy and Texture of Materials (Ministry of Education), School of Material Science and Engineering, Northeastern University, Shenyang 110819, China.
}%
\author{Ying Zhao}
\affiliation{%
Key Laboratory for Anisotropy and Texture of Materials (Ministry of Education), School of Material Science and Engineering, Northeastern University, Shenyang 110819, China.
}%
\author{Hao-Xuan Liu}
\affiliation{%
Key Laboratory for Anisotropy and Texture of Materials (Ministry of Education), School of Material Science and Engineering, Northeastern University, Shenyang 110819, China.
}%
\author{Shuchen Sun}
\email{sunsc@smm.neu.edu.cn}
\affiliation{%
School of Metallurgy, Northeastern University, Shenyang 110819, China.
}%
\author{Hongbo Yang}
\email{yanghongbo203@grirem.com}
\affiliation{%
Rare Earth Functional Materials (Xiong'an) Innovation Center Co., Ltd., Xiong'an 071700, China
}%
\author{Jiamin Zhong}
\affiliation{%
Rare Earth Functional Materials (Xiong'an) Innovation Center Co., Ltd., Xiong'an 071700, China
}%
\author{Ganfeng Tu}
\affiliation{%
School of Metallurgy, Northeastern University, Shenyang 110819, China.
}%
\author{Song Li}
\affiliation{%
Key Laboratory for Anisotropy and Texture of Materials (Ministry of Education), School of Material Science and Engineering, Northeastern University, Shenyang 110819, China.
}%
\author{Hai-Le Yan}
\email{yanhaile@mail.neu.edu.cn}
\affiliation{%
Key Laboratory for Anisotropy and Texture of Materials (Ministry of Education), School of Material Science and Engineering, Northeastern University, Shenyang 110819, China.
}%
\author{Liang Zuo}
\affiliation{%
Key Laboratory for Anisotropy and Texture of Materials (Ministry of Education), School of Material Science and Engineering, Northeastern University, Shenyang 110819, China.
}%

\begin{abstract}
Finite-temperature ductility-brittleness and electronic structures of Al$_3$Sc, Al$_2$Sc and AlSc are studied comparatively by first-principles calculations and \textit{ab-initio} molecular dynamics. Results show that Al$_3$Sc and Al$_2$Sc are inherently brittle at both ground state and finite temperatures. By contrast, AlSc possesses a significantly superior ductility evaluated from all Pugh’s, Pettifor’s and Poisson’s ductility-brittleness criteria. At ground state, AlSc meets the criteria of ductile according to Pugh’s and Poisson’s theories, while it is categorized as the brittle in the frame of Pettifor’s picture. With the increasing temperature, the ductility of all the studied compounds exhibits a noticeable improvement. In particular, as the temperature rises, the Cauchy pressure of AlSc undergoes a transition from negative to positive. Thus, at high temperatures (\textit{T} \textgreater 600 K), AlSc is unequivocally classified as the ductile from all criteria considered. In all Al$_3$Sc, Al$_2$Sc and AlSc, the Al-Al bond, originated from \textit{s}-\textit{p} and \textit{p}-\textit{p} orbital hybridizations, and the Al-Sc bond, dominated by \textit{p}-\textit{d} covalent hybridization, are the first and second strongest chemical bonds, respectively. To explain the difference in mechanical properties of the studied compounds, the mean bond strength (MBS) is evaluated. The weaker Al-Al bond in AlSc, leading to a smaller MBS, could be the origin for the softer elastic stiffness and superior intrinsic ductility. The longer length of the Al-Al bond in AlSc is responsible for its weaker bond strength. Furthermore, the enhanced metallicity of the Al-Al bond in AlSc would also contribute to its exceptional ductility.
\\ %
\end{abstract}

\maketitle

\section{Introduction}
Highly Sc-doped Al-Sc alloy target materials (with Sc of as high as 50 at. \%) comprised of several Al-Sc binary intermetallics, including Al$_3$Sc, Al$_2$Sc and AlSc, are crucial materials to fabricate (Al, Sc)N piezoelectric thin films that are widely utilized in microelectronic systems \cite{RN558,RN559,RN606,RN275}, such as radio frequency filters, piezoelectric actuators, and ultrasonic sensors. The microstructure uniformity of the Al-Sc target materials, including chemical composition, grain size, and crystallographic orientation, plays a pivotal role in the piezoelectric performance of the (Al, Sc)N films. Recently, it has been reported that the high-Sc Al-Sc alloys can be thermomechanical processing at elevated temperatures \cite{RN560,RN566,RN567}, which supplies an effective way to improve the microstructure uniformity of Al-Sc target materials. However, for different kinds of Al-Sc intermetallics contained in Al-Sc target materials, the information about the intrinsic brittleness-ductility, at both ground state and finite temperature, and the underlying mechanisms, is greatly limited. This situation severely impedes alloy composition design, thermomechanical process optimization, and ultimately the efficient fabrication of high-quality Al-Sc target materials.

To date, research on the binary Al-Sc intermetallics has primarily focused on Al$_3$Sc \cite{RN1,RN3,RN32,RN17,RN15,RN554,RN198,RN4,RN131,RN30,RN59}, due to the fact that the inclusion of Al$_3$Sc can substantially refine the microstructure and enhance the thermal stability of the conventional Al alloy structural materials. A. H. Reshak \textit{et al.} \cite{RN3} revealed that Al$_3$Sc exhibits a preference for the L1$_{2}$ structure over the D0$_{22}$ structure in thermodynamics. This preference is attributed to the smaller electronic states at the Fermi energy in the L1$_{2}$ structure. By Pugh’s (\textit{B}/\textit{G}) and Pettifor’s Cauchy pressure (\textit{C}$_{12}$–\textit{C}$_{44}$) ductility-brittleness criteria, R. Sharma \textit{et al.} \cite{RN1} suggested that the L1$_{2}$-type Al$_3$Sc is inherently brittle. D. Chen \textit{et al.} \cite{RN17} observed that the elastic moduli of Al$_3$Sc increase linearly with the increasing external hydrostatic pressure. In the range of 0 to 50 GPa, the brittle nature of Al$_3$Sc remains unchanged, while the elastic anisotropy becomes more pronounced. Furthermore, the finite-temperature elastic properties of Al$_3$Sc were investigated by R.-K. Pan \textit{et al.} using the quasi-harmonic approximation \cite{RN59}. With the elevated temperature, the elastic moduli of Al$_3$Sc exhibit a gentle descent trend but its brittle nature is unchanged. These observations are attributed to the slight weakening of the covalent bonding in Al$_3$Sc as temperature increases. Different from the situation of Al$_3$Sc, the studies on the intrinsic mechanical properties of Al$_2$Sc and AlSc are extremely limited. Only available results were reported by Ş. Uğur and coworkers \cite{RN547}. They found that, in contrast to the brittle nature of Al$_3$Sc, AlSc is inherently ductile in terms of Pugh’s criterion. These investigations significantly advance our understanding of the intrinsic ductility-brittleness of the Al-Sc binary intermetallics.

However, the intrinsic mechanical properties and the ductility-brittleness origins of different Al-Sc intermetallics remain elusive. In particular, there is a critical shortage of comparative research on intrinsic mechanical properties of different Al-Sc intermetallics, both at ground state and finite temperatures. Specifically, the following key issues remain unknown. \textit{\color{blue}i)} Are there additional stable Al-Sc intermetallics with Sc content less than 50 at. \% besides Al$_3$Sc, Al$_2$Sc, and AlSc? A recent study \cite{RN556} has reported that Al$_3$Sc$_{2}$, which is not depicted in the Al-Sc phase diagram, also exhibits a high thermodynamics stability. Therefore, conducting a systematic search for potential stable structures in the Al-Sc binary systems holds immense significance. \textit{\color{blue}ii)} Which type of Al-Sc intermetallic exhibits the greatest ductility? This information is of utmost importance for the composition design of the Al-Sc target materials. \textit{\color{blue}iii)} What are the temperature effects on the inherent ductility of different Al-Sc intermetallics? This knowledge is essential for the optimization of thermomechanical processes. \textit{\color{blue}iv)} What is the origin behind the varying ductility-brittleness exhibited by different kinds of Al-Sc intermetallics? Clarifying the mechanisms of the difference in ductility-brittleness among different Al-Sc intermetallics is crucial for the design of novel intermetallics with superior ductility.

To bridge these knowledge gaps, the stable phase searching, the ground-state and finite-temperature intrinsic mechanical properties, and the electronic structures of various Al-Sc intermetallics are studied systematically by first-principles calculations and \textit{ab-initio} molecular dynamics. First, the potential stable phases in the Al-Sc binary system are searched systematically using the variable-composition evolutionary structure search algorithm \cite{RN555,RN545,RN546} (Section \ref{sec:phasestability}). Second, the ground-state elastic moduli, the ductility-brittleness behaviors in terms of Pugh’s ratio \textit{B}/\textit{G}, Pettifor’s Cauchy pressure \textit{C}$_{12}$‒\textit{C}$_{44}$ and Poisson’s ratio \textit{v}, are investigated (Section \ref{sec:mechanicalground}). Third, the temperature dependences of elastic moduli and ductility-brittleness of different Al-Sc intermetallics are studied comparatively (Section \ref{sec:mechanicalfinitetem}). Lastly, the electronic structures, including band structure, density of states, charge density difference, electron localization function and crystal orbital Hamilton population, are analyzed (Section \ref{sec:electronicstructure}). The underlying mechanisms of various ductility-brittleness exhibited by different Al-Sc intermetallics are finally discussed (Section \ref{sec:discussion}).

\section{Methodology}
First-principles calculations were performed using density functional theory (DFT) implemented in the Vienna \textit{ab initio} Simulation Package (VASP 6.3) \cite{RN459,RN581}. The Perdew-Burke-Ernzerhof (PBE) parametrization of the generalized gradient approximation (GGA) was employed to describe the exchange-correlation function \cite{RN460,RN464}. The electron-ion interactions were described by the projector augmented wave (PAW) pseudopotential approach. The valence electron configurations of Al (3\textit{s}\textsuperscript{2}3\textit{p}\textsuperscript{1}) and Sc (3\textit{d}\textsuperscript{1}4\textit{s}\textsuperscript{2}) were adopted. A kinetic energy cutoff of 500 eV was adopted for wave-function expansion. The \textit{k}-point meshes with an interval of 0.03 $\times$ 2$\pi$ \AA$^{-1}$ for the Brillouin zone to ensure that all enthalpy calculations converged within 10$^{-5}$ eV. During structural relaxation, the force on each atom was relaxed to be less than 10$^{-2}$ eV/\AA. During the calculation of the electronic structure, the tetrahedron method with the Blöchl correction \cite{RN573} was used to integrate the Brillouin zone, and a denser \textit{k}-point mesh with an interval of 0.01 $\times$ 2$\pi$ \AA$^{-1}$ was adopted. Crystalline orbital Hamiltonian population (COHP) \cite{RN577,RN578} was calculated by using the Local Orbital Basis Suite Towards Electronic-Structure Reconstruction (LOBSTER) program \cite{RN570,RN569}. The charge analysis was performed using the Bader decomposition technique \cite{RN604}. The states of 3\textit{s} and 3\textit{p} for Al and 3\textit{d} and 4\textit{s} for Sc were taken as basis function sets. The phonon dispersion curves were calculated under the harmonic approximation by using the PHONOPY code \cite{RN629}.

The stable phase searching in the Al-Sc system was carried out using the generic evolutionary algorithm as implemented in the USPEX code \cite{RN555,RN545,RN546}. The variable composition mode was used to realize a high-efficient determination of different compositions. In the first generation, 200 structures were randomly created. Starting from the second generation, 120 structures were generated at each generation using six different operators: heredity (45\%), symmetric random generator (15\%), permutation (10\%), softmutation (10\%), transmutation (10\%), and lattice mutation (10\%). The maximum number of atoms in the unit cell was set to be 18. For each structure, to compromise speed and accuracy, five-step DFT calculations with increasing precision were adopted. To increase the efficiency of structure prediction, the known structures of Al$_3$Sc, Al$_2$Sc and AlSc were taken as the seeds. The search process continued until the 60th generation was reached. After the calculation, the stabilities of the predicted phases close to the convex hull were re-evaluated with higher precision.

The ground-state elastic constants were determined by computing the second-order derivatives of the total energy with respect to the position of the ions using a finite differences approach \cite{RN582}. The finite-temperature elastic constants were calculated by using \textit{ab-initio} molecular dynamics (AIMD) simulation. For Al$_3$Sc, Al$_2$Sc and AlSc, a 108-atom supercell with 3×3×3 L1$_{2}$-type unit cells, a 192-atom supercell with 2×2×2 C15-type unit cells, and a 128-atom supercell with 4×4×4 B2-type unit cells were adopted, respectively. During the calculations, the canonical ensemble (NVT) was employed, and the Nosé-Hoover thermostat \cite{RN571,RN572} was adopted to control the system temperature. The Verlet algorithm \cite{RN576} with a time step of 2 fs was used to solve Newton’s equations of motion. The finite-temperature elastic constants were determined with two consecutive steps. First, the equilibrium lattice constant at a certain temperature was determined. At a specific temperature, four independent 10-ps AIMD simulations were conducted. Each simulation started from an isotropically expanded ground state structure at a guessed expansion. At 300 K, the initial applied expansion for the four AIMD simulations were 0.3\%, 0.33\%, 0.4\%, and 0.43\%, respectively. After each simulation, the average internal pressure at each expansion can be obtained. With this information, the equilibrium lattice parameter at a certain temperature, \textit{i.e.}, the one with the average pressure near zero, can be determined. Second, at each temperature, five anisotropic strains (Eq. \ref{eq1}) with $\varepsilon$ ranging from –0.04 to 0.04 at a interval of 0.02 \cite{RN383} were separately applied.
\begin{equation}
\label{eq1}
\left(\begin{array}{l l l}{{\varepsilon}}&{{0}}&{{0}}\\ {{0}}&{{0}}&{{\frac{\varepsilon}{2}}}\\ {{0}}&{{\frac{\varepsilon}{2}}}&{{0}}\end{array}\right)    
\end{equation}
For each strained structure, a 10-ps AIMD simulation was performed to determine the stress tensor. By linear fitting between stress and strain with the following relations, the elastic constants at a specific temperature can be obtained.
\begin{equation}
\label{eq2}
\sigma_{1}\!=\!C_{11}\varepsilon,\, \sigma_{2}\!=\!{\sigma_{3}}\!=\!C_{12}\varepsilon,\, and\, \sigma_{4}\!=\!C_{44}\varepsilon
\end{equation}

\section{Results}
\subsection{Stable phase, crystal structure and phase stability}\label{sec:phasestability}
\begin{figure}[!ht]
    \centering
    \includegraphics[width=1\linewidth]{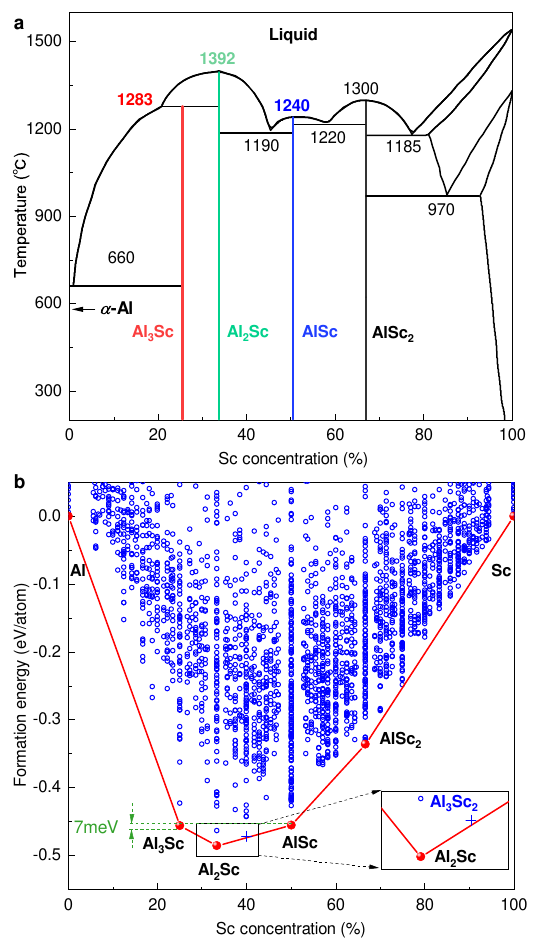}
    \caption{\textbf{Phase diagram and convex hull of the Al-Sc binary system.} (a) Al-Sc equilibrium phase diagram. The red, green, blue and black vertical lines indicate Al$_3$Sc, Al$_2$Sc, AlSc and AlSc$_{2}$ stable phase, respectively. (b) The calculated convex hull of the Al-Sc binary system. The blue circles indicate the formation energies and compositions of the predicted structures after each generation of the evolutionary algorithm search. The red line indicates the formation energies and compositions of the final thermodynamically stable Al-Sc phases, forming the convex hull. The inset is the enlarged region around Al$_3$Sc$_{2}$.}
    \label{fig:convexhull_phasediagram}
\end{figure}

The stable phases in the binary Al-Sc system are systematically explored by the variable-composition evolutionary structure search algorithm \cite{RN555,RN545,RN546}. The formation energies E$_{f}$ of the predicted Al-Sc phases at ambient pressure are shown in Fig. \ref{fig:convexhull_phasediagram}b. The structures with the lowest E$_{f}$ forming the convex hull are thought to be the ground-state stable phases. Clearly, the convex hull of the binary Al-Sc system is constituted by Al$_3$Sc with the Ni$_3$Al-type L1$_{2}$ structure (space group: Pm$\overline{3}$m, 221), Al$_2$Sc with the Cu$_{2}$Mg-type C15 structure (space group: Fd$\overline{3}$m, 227), AlSc with the CsCl-type B2 structure (space group: Pm$\overline{3}$m, 221) and AlSc$_{2}$ with the InNi$_{2}$-type B8$_{2}$ structure (space group: P6$_{3}$/mmc, 194). This result is in excellent agreement with the reported binary Al-Sc phase diagram \cite{RN275} (Fig. \ref{fig:convexhull_phasediagram}a). Note that E$_{f}$ of Al$_3$Sc$_{2}$ with space group Cmmm is slightly above the convex hull, indicated by the blue cross in the inset of Fig. \ref{fig:convexhull_phasediagram}b. This observation indicates that Al$_3$Sc$_{2}$ is not a thermodynamically stable phase, which is in good concordance with the investigation conducted by A. Bilić \textit{et al.} \cite{RN556}. Thus, in subsequent studies, we focus on Al$_3$Sc, Al$_2$Sc, and AlSc because these three intermetallics exist in the Al-Sc target materials with the Sc content not exceeding 50 at. \% \cite{RN65}.

\begin{figure*}
    \centering
    \includegraphics[width=1\linewidth]{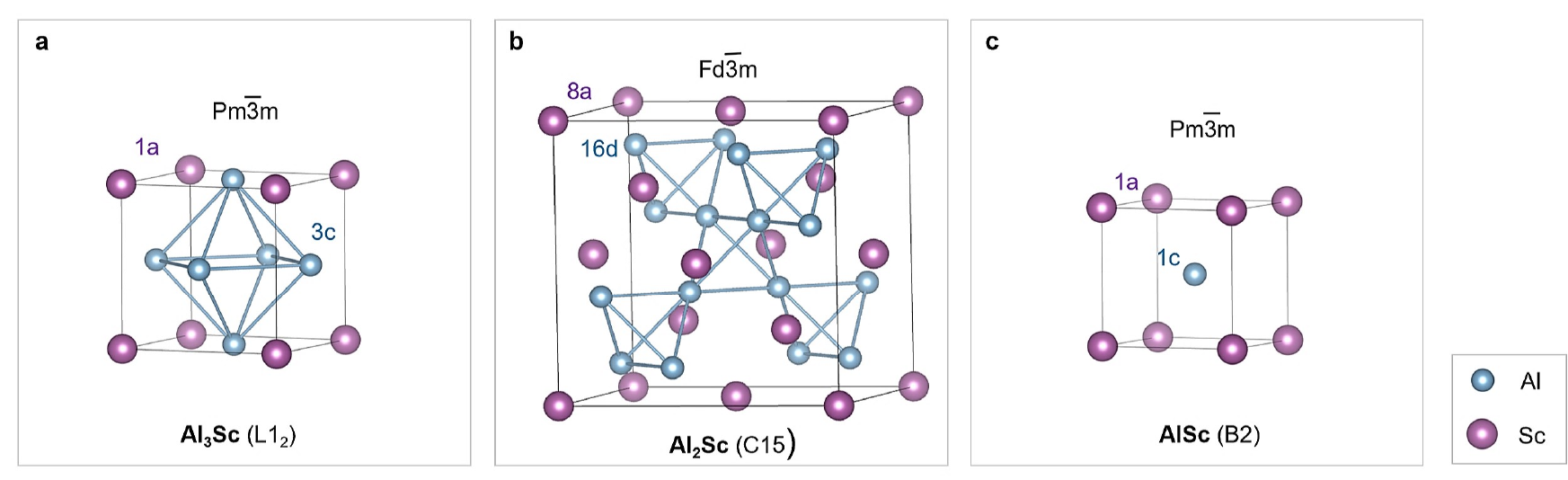}
    \caption{\textbf{Crystal structures of (a) Al$_3$Sc, (b) Al$_2$Sc and (c) AlSc.}}
    \label{fig:CrystalStructurel}
\end{figure*}

Phonon dispersion relation examinations show that the Al$_3$Sc with the L1$_{2}$ structure, the Al$_2$Sc with the C15 structure and the AlSc with the B2 structure are stable dynamically (\textit{see} Supplementary Materials Fig. $\color{red}{S1}$). Table \ref{tab:latticeparameter} lists the equilibrium lattice parameters a$_0$ of Al$_3$Sc, Al$_2$Sc and AlSc, along with the data extracted from the literature \cite{RN32,RN17,RN547,RN552,RN423,RN195,RN553,RN580,RN452,RN454,RN222}. The determined a$_0$ of Al$_3$Sc, Al$_2$Sc and AlSc are 4.103 Å, 7.572 Å and 3.378 Å, respectively, which are consistent with the experimental \cite{RN547,RN553,RN552} (with differences less than 1 \%) and theoretical reports \cite{RN32,RN17,RN195,RN452,RN454,RN222,RN423,RN580}. Fig. \ref{fig:CrystalStructurel}a, b and c display the crystal structure of Al$_3$Sc, Al$_2$Sc and AlSc, respectively. In the L1$_{2}$-type Al$_3$Sc, Al and Sc atoms occupy the \textit{3c} (0.5, 0, 0.5) and the \textit{1a} (0, 0, 0) sites, respectively. Each Al has 12 first-nearest neighboring atoms, comprising 8 Al and 4 Sc atoms. Similarly, each Sc also has 12 first-nearest neighboring atoms, which are all Al atoms. In the C15-type Al$_2$Sc, Sc and Al occupy the \textit{8a} (0, 0, 0) and the \textit{16d} (0.625, 0.625, 0.625) sites, respectively. Each Al atom has 6 first-nearest neighboring Al atoms. Four adjacent Al atoms form a tetrahedral structure. Each Sc atom, on the other hand, has no first-nearest neighboring atom. For the B2-type AlSc, Al and Sc atoms occupy the \textit{1a} (0, 0, 0) and \textit{1c} (0.5, 0.5, 0.5) sites, respectively. Each Al has 8 first-nearest neighboring Sc atoms, and each Sc has 8 first-nearest neighboring Al atoms. For phase stability, among Al$_3$Sc, Al$_2$Sc and AlSc, Al$_2$Sc exhibits the highest thermodynamic stability, evidenced by the smallest E$_{f}$ (Fig. \ref{fig:convexhull_phasediagram}b), which corresponds well to the highest melting point (1392°C) of Al$_2$Sc. E$_{f}$ of AlSc is comparable to that of Al$_3$Sc, with a slight difference of 7 meV/atom higher than Al$_3$Sc, which aligns with the slightly lower melting point of AlSc (1240°C) than Al$_3$Sc (1280°C).

\begin{table*}
\centering
\caption{\textbf{The ground-state equilibrium lattice parameters a$_0$ of  Al$_3$Sc, Al$_2$Sc and AlSc.}}
\label{tab:latticeparameter}
\begin{threeparttable}
\begin{tblr}{
  cells = {c},
  cell{1}{2} = {c=3}{},
  cell{1}{5} = {c=2}{},
  cell{1}{7} = {c=2}{},
  cell{2}{1} = {r=5}{},
  cell{2}{4} = {c=2}{},
  cell{3}{4} = {c=2}{},
  cell{4}{4} = {c=2}{},
  cell{5}{4} = {c=2}{},
  cell{6}{4} = {c=2}{},
  hline{1} = {-}{},
  hline{2} = {1}{},
  hline{2} = {2-8}{},
  hline{7} = {-}{},
}
Compounds & Al$_3$Sc                        &       &                             & Al$_2$Sc &                         & AlSc                         &       \\
a$_0$ (Å)    & Present                      & 4.103 & Present                     &       & 7.572                   & Present                      & 3.378 \\
          & Expt.\textsuperscript{ a}    & 4.101 & Expt.\textsuperscript{ e}   &       & 7.580                   & Expt.\textsuperscript{ i}    & 3.388 \\
          & GGA\textsuperscript{ b}      & 4.110 & PAW-GGA\textsuperscript{ f} &       & 7.581\textsuperscript{} & PW91-GGA\textsuperscript{ j} & 3.350 \\
          & PW91-GGA\textsuperscript{ c} & 4.109 & PAW-GGA\textsuperscript{ g} &       & 7.573                   & PAW-GGA\textsuperscript{ k}  & 3.372 \\
          & PBEsol\textsuperscript{ d}   & 4.098 & PBE-GGA\textsuperscript{ h} &       & 7.583                   & -                            & -     
\end{tblr}
\begin{tablenotes}
\footnotesize
\item \textsuperscript{a} Ref. \cite{RN552}; \textsuperscript{b} Ref. \cite{RN17}; \textsuperscript{c} Ref. \cite{RN423}; \textsuperscript{d} Ref. \cite{RN195}; \textsuperscript{e} Ref. \cite{RN553}; \textsuperscript{f} \cite{RN580}; \textsuperscript{g} Ref. \cite{RN452}; \textsuperscript{h} Ref. \cite{RN32}; \textsuperscript{i} Ref. \cite{RN547}; \textsuperscript{j} Ref. \cite{RN454}; \textsuperscript{k} Ref. \cite{RN222}.    
\end{tablenotes}
\end{threeparttable}
\end{table*}

\subsection{Intrinsic mechanical properties}
\subsubsection{Ground-state mechanical properties}\label{sec:mechanicalground}
To evaluate the ground-state mechanical properties, the elastic moduli of Al$_3$Sc, Al$_2$Sc and AlSc are calculated. Table \ref{tab:elasticmoduli} lists the independent elastic constants \textit{C}$_{ij}$ and the tetragonal shear modulus \textit{C}' (defined by (\textit{C}$_{11}$‒\textit{C}$_{12}$)/2) of Al$_3$Sc, Al$_2$Sc and AlSc. For Al$_3$Sc, \textit{C}$_{11}$, \textit{C}$_{12}$ and \textit{C}$_{44}$ are determined to be 188.8, 40.8 and 72.4 GPa, respectively, in good agreement with the report by R.-K. Pan and colleagues (\textit{i.e.}, 179.2, 40.6 and 71.0 GPa) \cite{RN59}. For all Al$_3$Sc, Al$_2$Sc and AlSc, their elastic constants satisfy the Born-Huang elastic stability criteria for cubic crystals \cite{RN579}, \textit{i.e.}, \textit{C}$_{11}$\textgreater0, \textit{C}$_{44}$\textgreater0, \textit{C}'\textgreater0 and (\textit{C}$_{11}$+2\textit{C}$_{12}$)\textgreater0, indicating that these intermetallics are elastically stable.

\begin{table}
\caption{\textbf{Independent ground-state elastic constants \textit{C}$_{11}$, \textit{C}$_{12}$, \textit{C}$_{44}$ and tetragonal shear modulus \textit{C}' of Al$_3$Sc, Al$_2$Sc and AlSc.}}
\label{tab:elasticmoduli}
\centering
\begin{threeparttable}
\begin{tabular*}{\linewidth}{cccccc}
\hline
\multicolumn{2}{c}{Compounds}      & \begin{tabular}[c]{@{}c@{}}\textit{C}$_{11}$\\ (GPa)\end{tabular} & \begin{tabular}[c]{@{}c@{}}\textit{C}$_{12}$\\ (GPa)\end{tabular} & \begin{tabular}[c]{@{}c@{}}\textit{C}$_{44}$\\ (GPa)\end{tabular} & \begin{tabular}[c]{@{}c@{}}\textit{C}'\\ (GPa)\end{tabular} \\
\hline
\multirow{2}{*}{Al$_3$Sc} & Present   & 188.8                                               & 40.8                                                & 72.4                                                & 74                                                 \\
                       & PBE-GGA \textsuperscript{ a} & 179.2                                               & 40.6                                                & 71                                                  & 69.3                                               \\
Al$_2$Sc                  & Present   & 67.5                                                & 40                                                  & 67.5                                                & 13.8                                               \\
AlSc                   & Present   & 96.6                                                & 73.2                                                & 91.5                                                & 11.7                                        \\ \hline
\end{tabular*}
\begin{tablenotes}
\footnotesize
\item \textsuperscript{a} Ref. \cite{RN59}
\end{tablenotes}
\end{threeparttable}
\end{table}

\begin{figure*}
    \centering
    \includegraphics[width=1\linewidth]{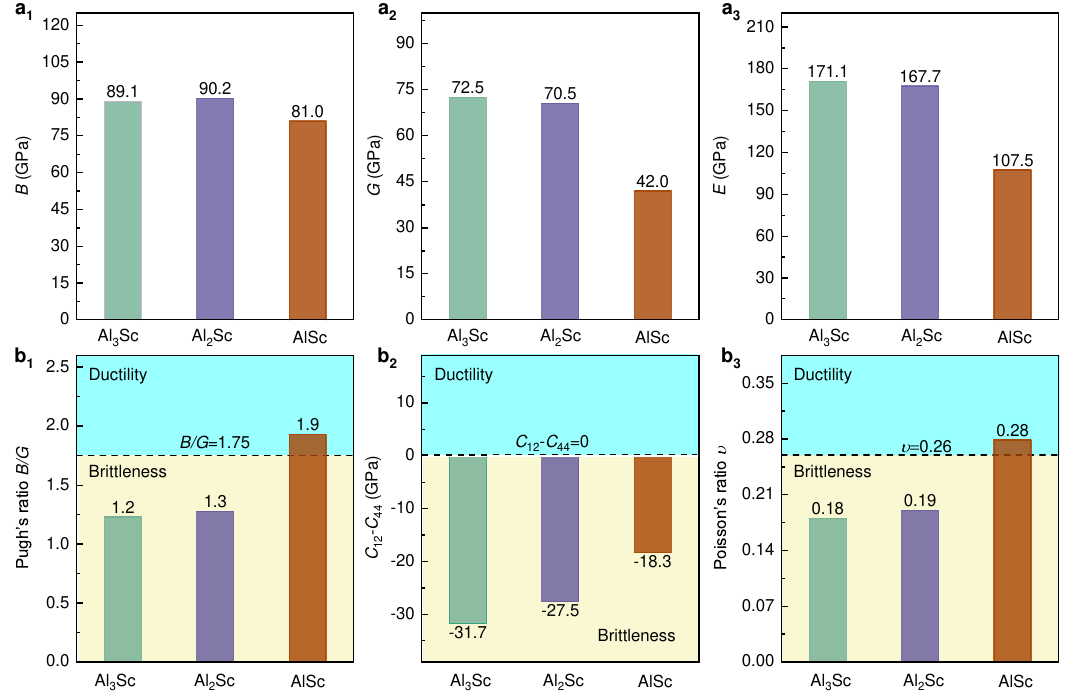}
    \caption{\textbf{Ground-state intrinsic mechanical properties.} (a$_{1}$) Bulk modulus \textit{B}; (a$_{2}$) Shear modulus \textit{G}; (a$_{3}$) Young’s modulus \textit{E}; (b$_{1}$) Pugh’s ratio \textit{B}/\textit{G}; (b$_{2}$) Cauchy pressure \textit{C}$_{12}$‒\textit{C}$_{44}$; (b$_{3}$) Poisson’s ratio \textit{v}.}
    \label{fig:ElasticModuli_0 K}
\end{figure*}

Fig. \ref{fig:ElasticModuli_0 K}a$_{1}$, a$_{2}$ and a$_{3}$ display isotropic bulk modulus \textit{B}, shear modulus \textit{G} and Young’s modulus \textit{E}, respectively. Herein, the Voigt-Ruess-Hill polycrystalline averaging algorithm\cite{RN396} is adopted (see details in Appendix \ref{appen:1}). It is evident that for all \textit{B}, \textit{G} and \textit{E}, the values of Al$_3$Sc are comparable to Al$_2$Sc, but both of them are obviously larger than that of AlSc. These results tell us that the elastic stiffnesses of Al$_3$Sc and Al$_2$Sc are higher than that of AlSc. Apart from elastic stiffness, the intrinsic ductility of Al$_3$Sc, Al$_2$Sc and AlSc are evaluated by using the three widely used ductility-brittleness criteria, \textit{i.e.}, Pugh’s ratio \textit{B}/\textit{G}, Pettifor’s Cauchy pressure \textit{C}$_{12}$‒\textit{C}$_{44}$ and Poisson’s ratio \textit{v}. In Fig. \ref{fig:ElasticModuli_0 K}b$_{1}$, b$_{2}$ and b$_{3}$, we display, respectively, \textit{B}/\textit{G}, \textit{C}$_{12}$‒\textit{C}$_{44}$ and $\nu$ for Al$_3$Sc, Al$_2$Sc and AlSc. For clarity, the critical minimum values for ductility, \textit{i.e.}, \textit{B}/\textit{G} = 1.75 in Pugh’s theory, \textit{C}$_{12}$‒\textit{C}$_{44}$ = 0 in Pettifor’s theory, and $\nu$ = 0.26 in Poisson’s theory \cite{RN398,RN458}, are plotted as dashed lines. In Al$_3$Sc, \textit{B}/\textit{G}, \textit{C}$_{12}$‒\textit{C}$_{44}$ and $\nu$ are determined to be 1.2, ‒31.7 GPa and 0.18, respectively. Based on all Pugh’s, Pettifor’s and Poisson’s criteria, Al$_3$Sc is classified as the intrinsically brittle, in line with the investigations of R. Sharma \textit{et al.} \cite{RN1} and D. Chen \textit{et al.} \cite{RN17}. In analogy to Al$_3$Sc, the C15-type Al$_2$Sc is also inherently brittle since all the values of \textit{B}/\textit{G} (1.3), \textit{C}$_{12}$‒\textit{C}$_{44}$ (‒27.5 GPa) and $\nu$ (0.19) are smaller than the critical minimum values of ductility.

As for the B2-type AlSc, the values of \textit{B}/\textit{G} (1.9) and $\nu$ (0.28) are higher than 1.75 and 0.26, respectively. Thus, AlSc is inherently ductile according to Pugh’s and Poisson’s criteria. Note that in terms of all Pugh’s, Pettifor’s and Poisson’s criteria, the values of Al$_3$Sc are comparable to Al$_2$Sc, while both of them are obviously smaller than those of AlSc. Therefore, it can be concluded that at ground state, the ductility of the B2-type AlSc is significantly superior to the L1$_{2}$-type Al$_3$Sc and the C15-type Al$_2$Sc. However, as seen in Fig. \ref{fig:ElasticModuli_0 K}b$_{2}$, the Cauchy pressure \textit{C}$_{12}$‒\textit{C}$_{44}$ of AlSc is negative (‒18.3 GPa). Thus, in the frame of Pettifor’s theory, AlSc should be considered to be brittle. This inconsistency of ductility-brittleness concluded by different criteria indicates that the inherent ductility of AlSc is not excellent enough. It may explain the challenges encountered when attempting severe plastic deformation in the high-Sc Al-Sc alloys at room temperature.

\subsubsection{Finite-temperature mechanical properties}\label{sec:mechanicalfinitetem}
\begin{figure*}
    \centering
    \includegraphics[width=1\linewidth]{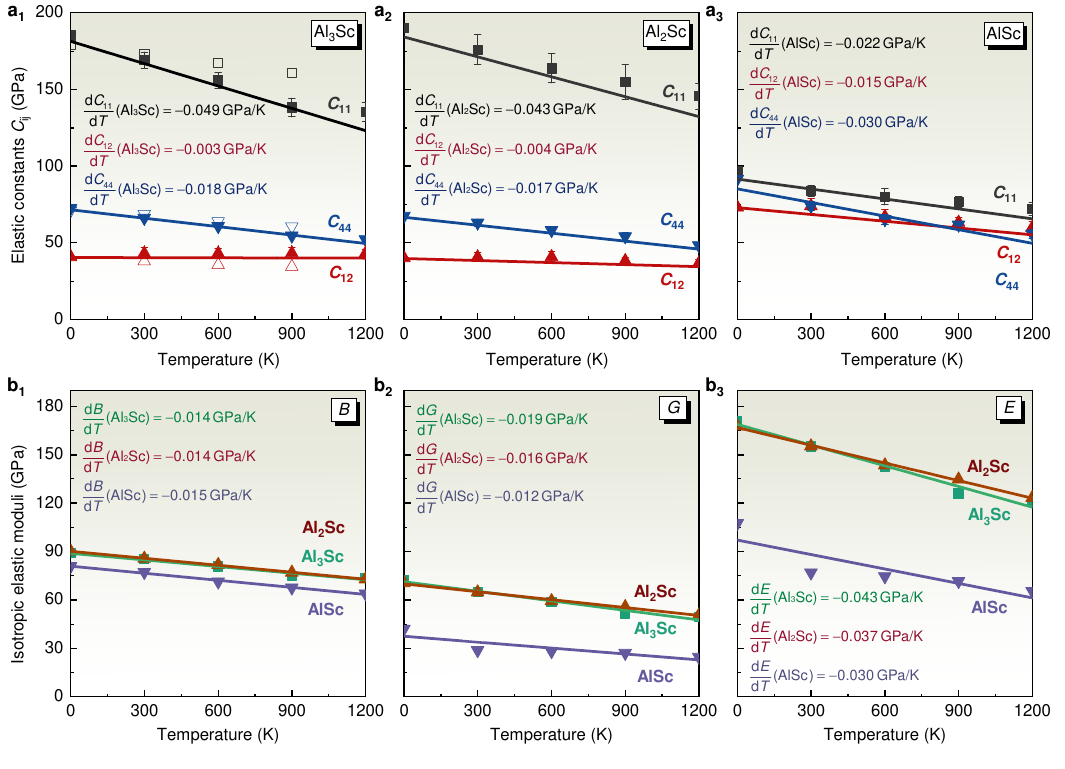}
    \caption{\textbf{Finite-temperature elastic moduli.} (a$_{1}$-a$_{3}$) Elastic constants of Al$_3$Sc, Al$_2$Sc and AlSc. The data represented by open symbols in (a) are calculated by quasi-harmonic approximation extracted from Ref. \cite{RN59}. (b$_{1}$-b$_{3}$) Isotropic bulk modulus \textit{B}, shear modulus \textit{G} and Young’s modulus \textit{E}.}
    \label{fig:ElasticModuli_FiniteTem}
\end{figure*}

To clarify the intrinsic mechanical properties at high temperatures, the finite-temperature crystal structures and elastic constants are calculated by the AIMD technique. Investigations show that for all Al$_3$Sc, Al$_2$Sc and AlSc, no structural transitions occur at the investigated temperatures (\textless 1200 K), in good agreement with the Al-Sc phase diagram \cite{RN275} (Fig. \ref{fig:convexhull_phasediagram}a). In Table \ref{tab:latticeparameter_finitetem}, we list the determined finite-temperature lattice parameters. Phonon dispersion relation examinations evidence that Al$_3$Sc with the L1$_{2}$ structure, Al$_2$Sc with the C15 structure and AlSc with B2 structure are still stable dynamically at the temperature up to 1200 K (\textit{see} Supplementary Materials Fig. $\color{red}{S2}$). Fig. \ref{fig:ElasticModuli_FiniteTem}a$_{1}$, a$_{2}$ and a$_{3}$ show the temperature dependences of independent elastic constants \textit{C}$_{ij}$ for Al$_3$Sc, Al$_2$Sc and AlSc, respectively. As a reference, the finite-temperature \textit{C}$_{ij}$ of Al$_3$Sc, calculated by the quasi-harmonic approximation \cite{RN59}, is also included in Fig. \ref{fig:ElasticModuli_FiniteTem}a$_{1}$ (open symbols). We see that the determined temperature dependences of \textit{C}$_{12}$, \textit{C}$_{11}$ and \textit{C}$_{44}$ in Al$_3$Sc by AIMD are in reasonable agreement with those calculated by the quasi-harmonic approximation \cite{RN59}.

\begin{table}[ht]
\centering
\caption{\textbf{Finite-temperatures equilibrium lattice parameters a$_0$ of Al$_3$Sc, Al$_2$Sc and AlSc}}
\label{tab:latticeparameter_finitetem}
\begin{tabular}{ccccc}
\cline{1-4}
\multirow{2}{*}{Temperature (K)} & \multicolumn{3}{c}{Lattice constant (Å)} &  \\ \cline{2-4}
                                 & Al$_3$Sc        & Al$_2$Sc       & AlSc        &  \\ \cline{1-4}
0                                & 4.103        & 7.572       & 3.378       &  \\
300                              & 4.119        & 7.599       & 3.385       &  \\
600                              & 4.135        & 7.625       & 3.398       &  \\
900                              & 4.152        & 7.653       & 3.411       &  \\
1200                             & 4.169        & 7.685       & 3.426       &  \\ \cline{1-4}
\end{tabular}
\end{table}

For all Al$_3$Sc, Al$_2$Sc and AlSc, there is a noticeable decrease in \textit{C}$_{11}$ and \textit{C}$_{44}$ as the temperature increases. The temperature-induced variation in elastic constants is majorly attributed to the expansion of lattice volume with the increasing temperature (Table \ref{tab:latticeparameter_finitetem}), which typically weakens the bonding strength between the constituent elements and further results in a decrease in elastic constants. For Al$_3$Sc and Al$_2$Sc, the decrease rate of \textit{C}$_{11}$ is significantly larger than that of \textit{C}$_{44}$. For instance, in Al$_3$Sc, the temperature sensitivity coefficient of \textit{C}$_{11}$ (\textit{i.e.}, d\textit{C}$_{11}$/dT) is ‒0.049 GPa/K, which is around three times larger than that of \textit{C}$_{44}$ (d\textit{C}$_{44}$/dT = ‒0.018 GPa/K). Nevertheless, in the case of AlSc, \textit{C}$_{44}$ exhibits a larger decrease rate (‒0.030 GPa/K) compared to \textit{C}$_{11}$ (‒0.022 GPa/K) (Fig. \ref{fig:ElasticModuli_FiniteTem}a$_{3}$). Different from \textit{C}$_{11}$ and \textit{C}$_{44}$, \textit{C}$_{12}$ shows a much weaker sensitivity on temperature, particularly for Al$_3$Sc (d\textit{C}$_{12}$/dT = ‒0.003 GPa/K) and Al$_2$Sc (d\textit{C}$_{12}$/dT = ‒0.004 GPa/K). It is worth noting that in the cases of Al$_3$Sc and Al$_2$Sc, \textit{C}$_{44}$ consistently remains larger than \textit{C}$_{12}$ even for the temperature up to 1200 K, whereas for AlSc there is a reversal in the relative sizes of \textit{C}$_{12}$ and \textit{C}$_{44}$ (Fig. \ref{fig:ElasticModuli_FiniteTem}a$_{3}$). As is known, the difference between \textit{C}$_{12}$ and \textit{C}$_{44}$, \textit{i.e.}, Cauchy pressure, is a crucial parameter that characterizes intrinsic ductility in Pettifor’s theory. The reversal in the relative magnitudes of \textit{C}$_{12}$ and \textit{C}$_{44}$ suggests a great improvement of ductility under the assistance of temperature, which is detailedly discussed later.

Fig. \ref{fig:ElasticModuli_FiniteTem}b$_{1}$, b$_{2}$ and b$_{3}$ display the temperature dependences for isotropic bulk modulus \textit{B}, shear modulus \textit{G} and Young’s modulus \textit{E}, respectively. With the increasing temperature, all values of \textit{B}, \textit{G}, and \textit{E} exhibit a linearly decreasing tendency. At the examined temperatures, Al$_3$Sc and Al$_2$Sc possess similar values for \textit{B}, \textit{G}, and \textit{E}, which are significantly larger compared to those of AlSc. For \textit{B}, the temperature sensitivity coefficients of Al$_3$Sc, Al$_2$Sc, and AlSc are nearly identical, with values around ‒0.014 GPa/K (Fig. \ref{fig:ElasticModuli_FiniteTem}b$_{1}$). Nevertheless, noticeable differences are observed for the temperature sensitivities of \textit{G} and \textit{E}. For \textit{G}, the temperature sensitivity coefficients of Al$_3$Sc, Al$_2$Sc and AlSc are ‒0.019, ‒0.016 and ‒0.012 GPa/K, respectively, while these three values for \textit{E} are ‒0.043, ‒0.037 and ‒0.030 GPa/K, respectively. Clearly, for both \textit{G} and \textit{E}, the temperature sensitivities in Al$_3$Sc are the highest, while those in AlSc are the lowest. Moreover, it is worth noting that for different elastic moduli, the temperature sensitivities exhibit significant differences. Among  \textit{B}, \textit{E} and \textit{G}, \textit{E} exhibits the highest temperature sensitivity. For all the studied compounds, the temperature sensitivity coefficients of \textit{E} are around ‒0.035 GPa/K, which is approximately twice larger than those of \textit{B} and \textit{G} (about ‒0.015 GPa/K).

\begin{figure}[!ht]
    \centering
    \includegraphics[width=1\linewidth]{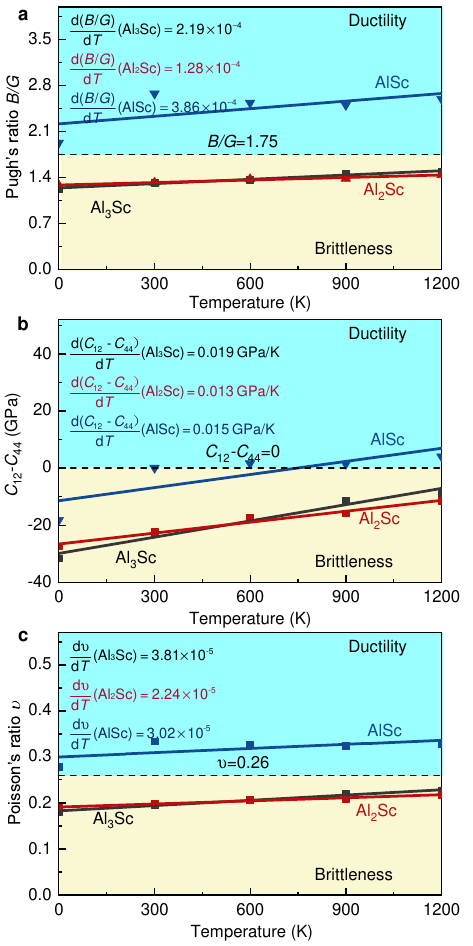}
    \caption{\textbf{Temperature dependence of intrinsic ductility-brittleness of Al$_3$Sc, Al$_2$Sc and AlSc.} (a) Pugh’s ratio \textit{B}/\textit{G}; (b) Cauchy pressure \textit{C}$_{12}$‒\textit{C}$_{44}$; (c) Poisson’s ratio \textit{v}.}
    \label{fig:ductility_finitetemsl}
\end{figure}

Fig. \ref{fig:ductility_finitetemsl}a, b and c display the temperature dependences of  \textit{B}/\textit{G}, \textit{C}$_{12}$‒\textit{C}$_{44}$ and $\nu$, respectively. For Al$_3$Sc, Al$_2$Sc, and AlSc, all \textit{B}/\textit{G}, \textit{C}$_{12}$‒\textit{C}$_{44}$ and $\nu$ exhibit a monotonically increasing tendency with the elevated temperature. This result suggests that the intrinsic ductility of all compounds can be effectively improved under the assistance of temperature, which aligns well with the experimental observations of the improved plasticity deformation ability of Al-Sc alloys at high temperatures \cite{RN560,RN566,RN567}. At the investigated temperatures, \textit{B}/\textit{G}, \textit{C}$_{12}$‒\textit{C}$_{44}$ and $\nu$ of AlSc are always larger than those of Al$_3$Sc and Al$_2$Sc, indicating that AlSc still exhibits a noticeably superior intrinsic ductility compared to Al$_3$Sc and Al$_2$Sc at high temperatures. As described earlier, at ground state, AlSc is classified as the ductile based on Pugh’s and Poisson’s theories but as the brittle according to Pettifor’s criterion. However, with the increasing temperature, the Cauchy pressure of AlSc evolves from negative to positive (Fig. \ref{fig:ductility_finitetemsl}b). Consequently, at high temperatures (T \textgreater around 600 K), according to all Pugh’s, Pettifor’s, and Poisson’s theories, the B2-type AlSc should be treated as the inherently ductile. This may be the reason for the high plasticity deformation ability of high-Sc Al-Sc target materials at high temperatures \cite{RN560,RN566,RN567}. These findings could serve as a crucial foundation for the composition design and the thermomechanical processing optimization of Al-Sc target materials. 

Furthermore, we note that as the temperature increases, the relative ductility of Al$_3$Sc and Al$_2$Sc changes. At ground state, the values of \textit{B}/\textit{G}, \textit{C}$_{12}$‒\textit{C}$_{44}$ and $\nu$ for Al$_3$Sc are slightly smaller than those of Al$_2$Sc. It indicates that Al$_3$Sc exhibits an inferior ductility compared to Al$_2$Sc, although the differences are not significant. However, the relative magnitudes of \textit{B}/\textit{G}, \textit{C}$_{12}$‒\textit{C}$_{44}$ and $\nu$ for Al$_3$Sc and Al$_2$Sc undergo a shift when the temperature exceeds around 600 K (Fig. \ref{fig:ductility_finitetemsl}). This result implies that Al$_3$Sc possesses superior high-temperature ductility compared to Al$_2$Sc. This change is ascribed to the higher temperature sensitivities of \textit{B}/\textit{G}, \textit{C}$_{12}$‒\textit{C}$_{44}$ and $\nu$ of Al$_3$Sc. For Al$_3$Sc, the temperature sensitivity coefficients of \textit{B}/\textit{G}, \textit{C}$_{12}$‒\textit{C}$_{44}$ and $\nu$ are fitting to be 2.19 $\times$ 10\textsuperscript{‒4}, 0.019 GPa/K and 3.81 $\times$ 10\textsuperscript{‒5}, respectively, which are significantly larger than those of Al$_2$Sc, \textit{i.e.}, 1.28 $\times$ 10\textsuperscript{‒4}, 0.013 GPa/K and 2.24 $\times$ 10\textsuperscript{‒5}. Analysis shows that the higher temperature sensitivity of Al$_3$Sc could be attributed to the higher thermal expansion coefficient of this compound. With the finite-temperature equilibrium lattice parameters determined by AIMD simulation (Table \ref{tab:latticeparameter_finitetem}), the linear thermal expansion coefficient of Al$_3$Sc is found to be 1.34 $\times$ 10\textsuperscript{‒5} K\textsuperscript{‒1}, which is larger than that of Al$_2$Sc (1.23 $\times$ 10\textsuperscript{‒5} K\textsuperscript{‒1}) as well as AlSc (1.20 $\times$ 10\textsuperscript{‒5} K\textsuperscript{‒1}). However, it is important to note that despite a great improvement in the ductilities of Al$_3$Sc and Al$_2$Sc with increasing temperature, both these two compounds retain their brittle nature even at temperatures up to 1200 K.

\subsection{Electronic structures}\label{sec:electronicstructure}
\subsubsection{Band structure and DOS}
\begin{figure*}[ht]
    \centering
    \includegraphics[width=1\linewidth]{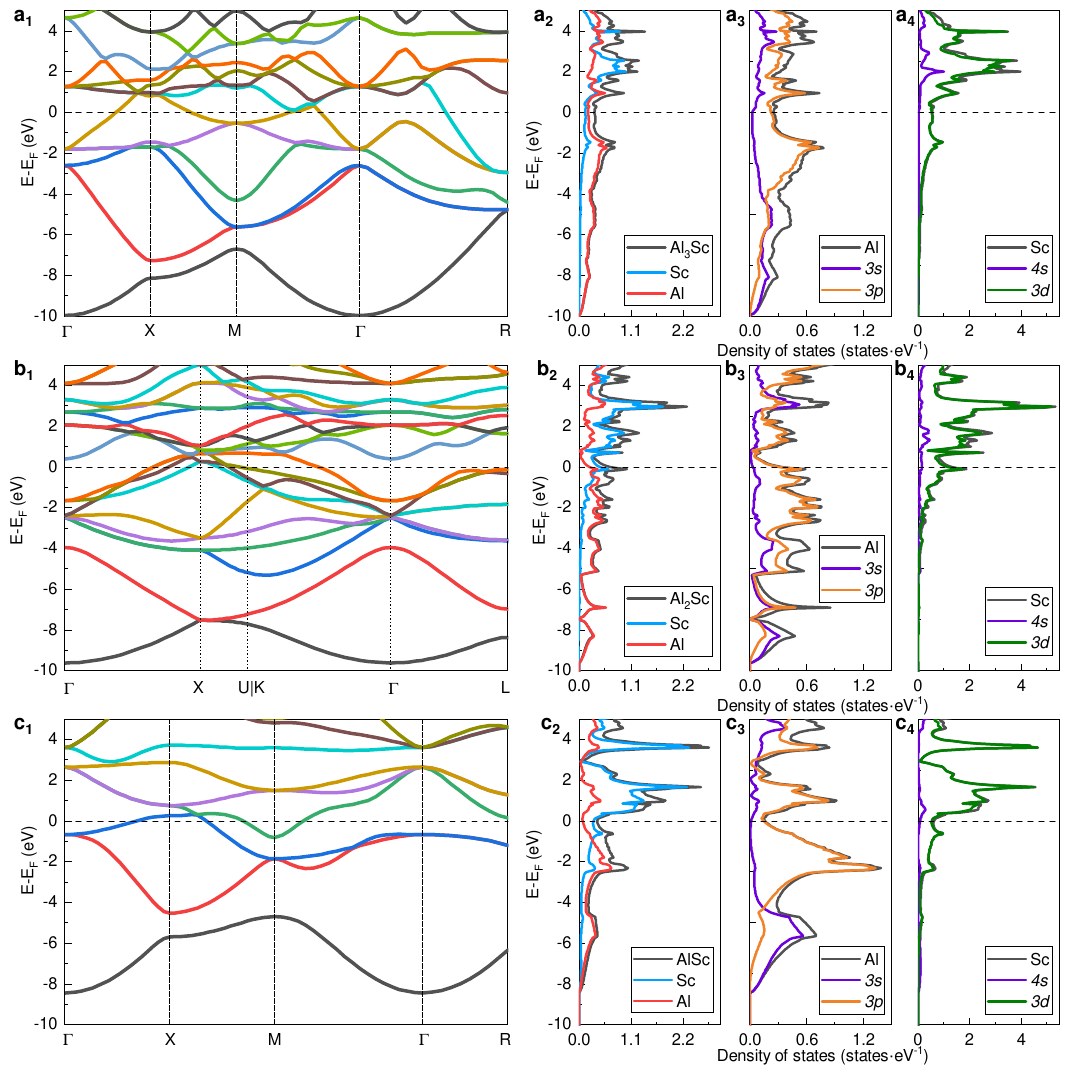}
    \caption{\textbf{Electronic band structures and density of states (DOSs) of (a$_{1}$-a$_{4}$ )Al$_3$Sc, (b$_{1}$-b$_{4}$) Al$_2$Sc and (c$_{1}$-c$_{4}$) AlSc.} a$_{1}$-c$_{1}$, a$_{2}$-c$_{2}$, a$_{3}$-c$_{3}$ and a$_{4}$-c$_{4}$ are electronic band structure, total DOS, partial DOS of Al, and partial DOS of Sc, respectively. The horizontal dashed line represents the Fermi energy \textit{E}$_F$.}
    \label{fig:bs}
\end{figure*}

\begin{figure}[!ht]
    \centering
    \includegraphics[width=1\linewidth]{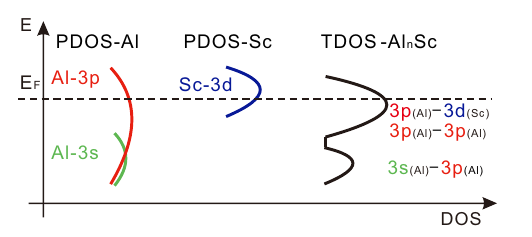}
    \caption{\textbf{Illustration of the DOS structure of the studied binary Al-Sc compounds.}}
    \label{fig:illustrationDosl}
\end{figure}

To understand the origin of different intrinsic mechanical properties of Al$_3$Sc, Al$_2$Sc, and AlSc, a thorough investigation of their electron structures is conducted, including band structure, density of states (DOS), electron localization function (ELF), charge density difference (CDD), Bader charge and crystalline orbital Hamiltonian population (COHP). Fig. \ref{fig:bs}a, b and c display the electronic band structures and the total and partial DOSs for Al$_3$Sc, Al$_2$Sc and AlSc, respectively. Analyses reveal that despite the significant differences in composition and structure, the electronic band and DOS structures of Al$_3$Sc, Al$_2$Sc and AlSc exhibit some common features, as depicted in Fig. \ref{fig:illustrationDosl} and detailed below. In all Al$_3$Sc, Al$_2$Sc, and AlSc, the 3\textit{s} electrons of Al are primarily distributed within the energy range of ‒9 to ‒3 eV; by contrast, the 3\textit{p} electrons of Al appear at a broad region with energies above ‒9 eV (Fig. \ref{fig:bs}a$_{3}$-c$_{3}$). As for the Sc atom, its 3\textit{d} electron predominantly occupies the energies above ‒3 eV, while below the Fermi level the 4s electrons are limited (Fig. \ref{fig:bs}a$_{4}$-c$_{4}$). For all the examined compounds, in the low-energy region, a strong \textit{s}-\textit{p} hybridization between Al and Al is observed (Fig. \ref{fig:bs}a$_{2}$-c$_{2}$). When the energy approaches the Fermi level, the \textit{p}-\textit{p} hybridization between Al-Al and the \textit{p}-\textit{d} orbital interaction between Al and Sc becomes dominant.

\subsubsection{ELF, CDD and Bader charge}
\begin{figure*}
    \centering
    \includegraphics[width=1\linewidth]{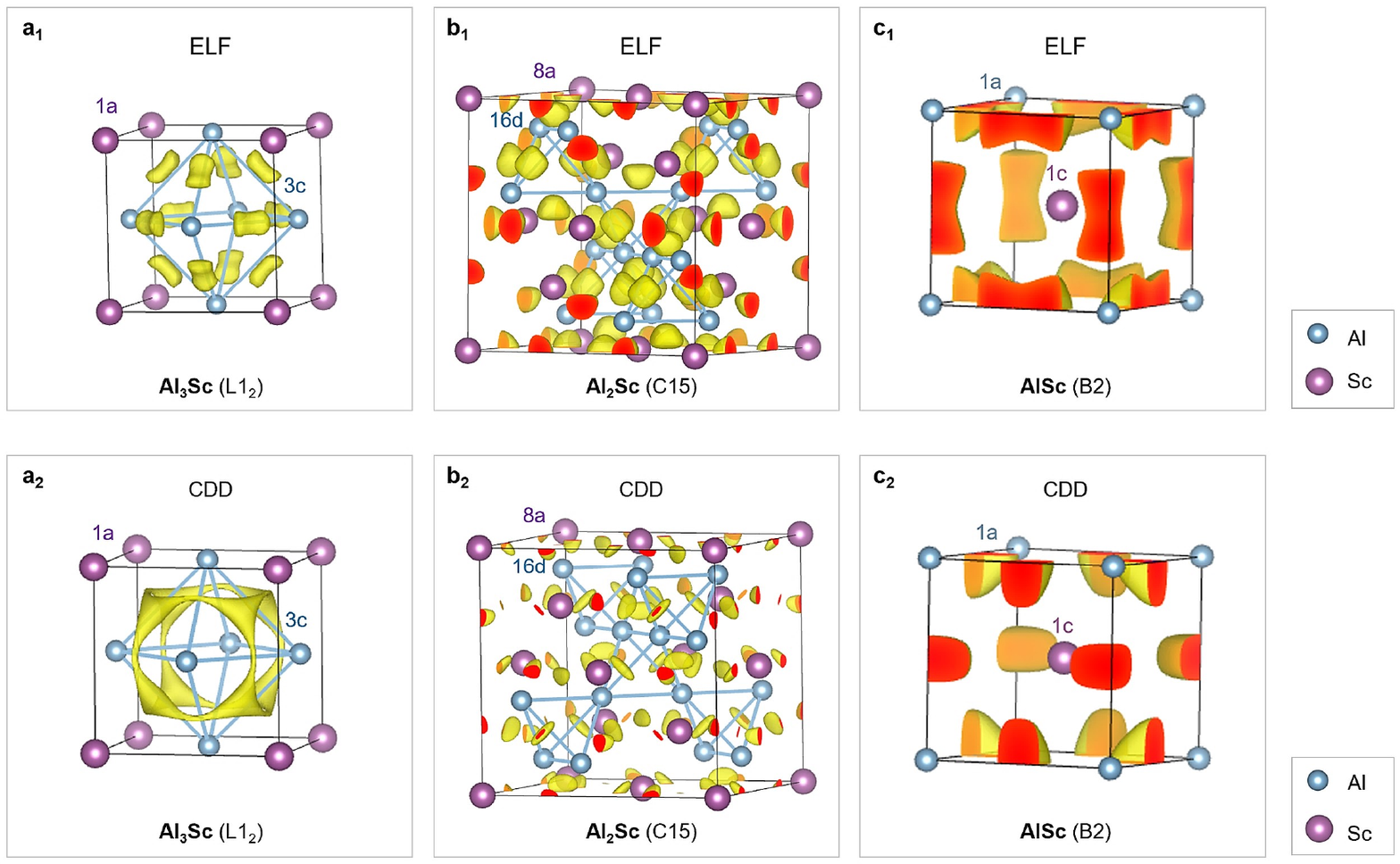}
    \caption{\textbf{Isosurfaces of electron localization function (ELF) and charge distribution difference (CDD).} (a$_{1}$-a$_{2}$) Al$_3$Sc; (b$_{1}$-b$_{2}$) Al$_2$Sc; (c$_{1}$-c$_{2}$) AlSc. The isosurface values for ELF (a$_{1}$, b$_{1}$ and c$_{1}$) and CDD (a$_{2}$, b$_{2}$ and c$_{2}$) are set to be 0.7 and 0.00505, respectively, to signify the maxima.}
    \label{fig:elf3d}
\end{figure*}

We shift our focus on electron structure from the reciprocal space to the real space. In Fig. \ref{fig:elf3d}a, b and c, we display the three-dimensional (3D) isosurfaces of ELF and CDD for Al$_3$Sc, Al$_2$Sc and AlSc, respectively. The isosurface values of ELF and CDD are carefully selected to signify the maxima. It is observed that in all Al$_3$Sc, Al$_2$Sc, and AlSc, there are prominent highly localized electrons and electron accumulation between the neighboring Al and Al atoms, indicating a strong chemical interaction of the Al-Al pairs. The strong interaction between Al-Al pairs majorly arises from the hybridizations of 3\textit{s}-3\textit{p} and 3\textit{p}-3\textit{p} orbital electrons, as revealed in the DOS analyses (Fig. \ref{fig:bs}). For AlSc, the maximum of ELF appears roughly at the center of the Al-Al pair (Fig. \ref{fig:elf3d}c$_{1}$). Nevertheless, in the cases of both Al$_3$Sc (Fig. \ref{fig:elf3d}a$_{1}$) and Al$_2$Sc (Fig. \ref{fig:elf3d}b$_{1}$), an obvious deviation of the ELF maximum from the center of the Al-Al pair is observed.

\begin{figure*}
    \centering
    \includegraphics[width=1\linewidth]{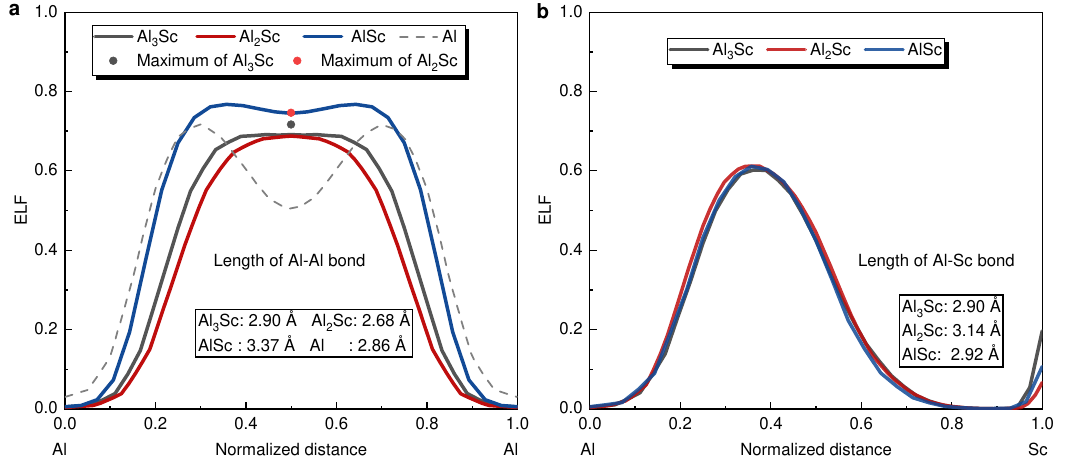}
    \caption{\textbf{ELF profiles along (a) Al-Al and (b) Al-Sc.} For easy comparison, the lengths of Al-Al and Al-Sc pairs in different compounds are normalized.}
    \label{fig:elf2d}
\end{figure*}

Although visualizing the ELF maximum in the 3D representation helps identify the strongest interaction between the constituent atoms, it may lose the information of relatively weaker interactions. To address this limitation, the line profile analyses of ELF between neighboring atoms are conducted. In Fig. \ref{fig:elf2d}a and b, we display the ELF distributions along the adjacent Al-Al and Al-Sc pair, respectively. Herein, to realize a direct comparison, the lengths of Al-Al and Al-Sc pairs in different compounds are normalized. Besides, to better illustrate the electron localization around the Al-Al pair, the maximum of ELF in Al$_3$Sc and Al$_2$Sc that slightly deviates from the Al-Al connection line are also displayed in Fig. \ref{fig:elf2d}a. We see that in terms of the maximum values of ELF around the Al-Al pair, Al$_3$Sc, Al$_2$Sc and AlSc are very close to each other. However, the ELF profile of the Al-Al pair in AlSc is much wider than those of Al$_3$Sc and Al$_2$Sc. The full width at half maximum (FWHM) of the ELF profiles of AlSc is measured to be 0.65, which is significantly longer than those of Al$_3$Sc (0.56) and Al$_2$Sc (0.51). In fact, the width of the ELF profile of the Al-Al pair in AlSc is compared to that of pure Al with excellent ductility, as indicated by the dashed line in Fig. \ref{fig:elf2d}a. Thus, in terms of the metallicity of the Al-Al bond, AlSc should be superior to Al$_3$Sc and Al$_2$Sc.

From Fig. \ref{fig:elf2d}b, it is observed that there exists pronounced localized electrons between Al and Sc, which indicates a non-negligible covalent interaction between the adjacent Al and Sc. From the DOS result (Fig. \ref{fig:bs}), it is clear that this interaction is primarily attributed to the \textit{p}-\textit{d} orbital interaction between the 3\textit{p} electrons of Al and the 3\textit{d} orbitals of Sc. Remarkably, despite the significant differences between the examined compounds, the ELF profiles along Al and Sc exhibit a remarkable overlap, implying a similarity in the bonding characteristics along Al-Sc. Furthermore, we note that the position of the ELF maximum deviates from the center of Al-Sc pair and shifts towards the Al atom (Fig. \ref{fig:elf2d}). This is attributed to the larger electronegativity of Al (1.61) compared to Sc (1.36), indicating the polarity nature of the Al-Sc bond.

In Table \ref{tab:bader}, we list the transferred charge between Al and Sc during bonding following Bader’s idea of the partition on space charge \cite{RN604}. Compared to neutral atoms, the charge amount of Al increases while that of Sc decreases, which is consistent with the polarity of the Al-Sc bond. In Al$_3$Sc, Al$_2$Sc, and AlSc, the transferred electrons from a Sc atom are approximately the same (around 1 electron), which are evenly distributed among the surrounding Al atoms. In AlSc, a single Al atom gets the most electrons (1.03) while receiving the least electrons (0.34) in Al$_3$Sc. This difference is attributed to the different Al content in the studied three compounds, with AlSc having the lowest (50 at. \%) and Al$_3$Sc having the highest (75 at. \%) one.

\begin{table}[!ht]
\caption{\textbf{Charge transfer during bonding of Al$_3$Sc, Al$_2$Sc and AlSc.}}
\label{tab:bader}
\begin{tabular*}{\linewidth}{ccccccc}
\hline
\multirow{2}{*}{Compounds} & \multicolumn{2}{c}{\begin{tabular}[c]{@{}c@{}}Bader charge\\ in neutra\\ atoms\end{tabular}} & \multicolumn{2}{c}{\begin{tabular}[c]{@{}c@{}}Bader charge\\ in compounds\end{tabular}} & \multicolumn{2}{c}{\begin{tabular}[c]{@{}c@{}}Transferred\\ charge\\ during bonding\end{tabular}} \\ \cline{2-7} 
                           & Al                                            & Sc                                           & Al                                         & Sc                                         & Al                                              & Sc                                              \\ \hline
Al$_3$Sc                       & 3                                             & 3                                            & 3.34                                       & 1.97                                       & 0.34                                            & ‒1.03                                           \\
Al$_2$Sc                      & 3                                             & 3                                            & 3.56                                       & 1.89                                       & 0.56                                            & ‒1.11                                           \\
AlSc                       & 3                                             & 3                                            & 4.03                                       & 1.97                                       & 1.03                                            & ‒1.03                                           \\ \hline
\end{tabular*}
\end{table}

\subsubsection{Chemical bonding}
\begin{figure}[!ht]
    \centering
    \includegraphics[width=1\linewidth]{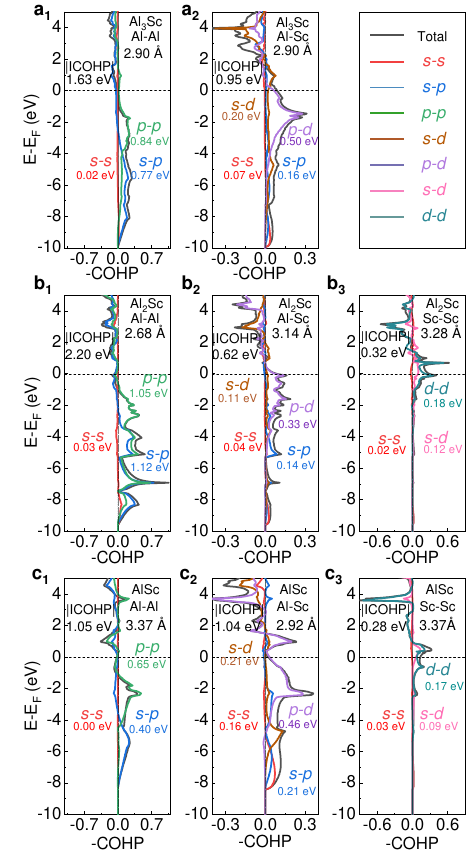}
    \caption{\textbf{COHP and orbital-resolved COHP curves of chemical bonds with lengths of less than 3.5 Å.} First neighboring (a$_1$) Al-Al and (a$_2$) Al-Sc bond in Al$_3$Sc. (b$_1$) First neighboring Al-Al, (b$_2$) second neighboring Al-Sc, (b$_3$) third neighboring Sc-Sc bonds in Al$_2$Sc. (c$_1$) Second neighboring Al-Al, (b$_2$) first neighboring Al-Sc, (b$_3$) second neighboring Sc-Sc bonds in AlSc. For each bond, the bond length, the total and orbital-resolved \textbar{}ICOHP\textbar{} values are also displayed.}
    \label{fig:cohp}
\end{figure}

To gain a comprehensive understanding of chemical bonding of Al$_3$Sc, Al$_2$Sc and AlSc, the crystalline orbital Hamiltonian population (COHP) method \cite{RN577,RN578} is adopted. Herein, we focus on the chemical bonds with lengths less than 3.5 Å because the strengths of bonds with longer lengths are very weak. Fig. \ref{fig:cohp}a, b and c display the total and orbital-resolved COHP curves of chemical bonds in Al$_3$Sc, Al$_2$Sc and AlSc, respectively. In Table \ref{tab:ICOHP} and Table \ref{tab:pCOHP}, we provide a summary of chemical bond type, bond length, bond number in a unit cell, and the absolute values of the integration of COHP (\textbar{}ICOHP\textbar{}) and the orbital-resolved \textbar{}ICOHP\textbar{}. For the L1$_{2}$-type Al$_3$Sc, only the first neighboring Al-Al (Fig. \ref{fig:cohp}a$_{1}$) and Al-Sc (Fig. \ref{fig:cohp}a$_{2}$) bonds with a length of 2.90 Å are included. In the case of C15-type Al$_2$Sc, the first neighboring Al-Al (Fig. \ref{fig:cohp}b$_{1}$) with a length of 2.68 Å, the second neighboring Al-Sc (Fig. \ref{fig:cohp}b$_{2}$) with a length of 3.14 Å, and the third neighboring Sc-Sc (Fig. \ref{fig:cohp}b$_{3}$) with a length of 3.28 Å bonds are satisfied. For the B2-type AlSc, the first neighboring Al-Sc (Fig. \ref{fig:cohp}c$_{2}$) with a length of 2.92 Å and the second neighboring Al-Al (Fig. \ref{fig:cohp}c$_{1}$) and Sc-Sc (Fig. \ref{fig:cohp}c$_{3}$) bonds with a length of 3.37 Å are covered. 

For all Al$_3$Sc, Al$_2$Sc, and AlSc, \textbar{}ICOHP\textbar{} of the Al-Al bond possesses the largest value among all chemical bonds in each compound. This result suggests that in all examined compounds, the Al-Al bond is the strongest chemical bond, in good accordance with the observations of ELF and CDD (Fig. \ref{fig:elf3d}). From the orbital-resolved COHP curves and \textbar{}ICOHP\textbar{} (Fig. \ref{fig:cohp}a$_{1}$-c$_{1}$ and Table \ref{tab:pCOHP}), it is clear that the strength of Al-Al bond majorly arises from \textit{s}-\textit{p} hybridization at the low-energy region and the \textit{p}-\textit{p} hybridization near the Fermi level, in agreement with the DOS analyses (Fig. \ref{fig:illustrationDosl}). Among Al$_3$Sc, Al$_2$Sc, and AlSc, the Al-Al bonds in Al$_3$Sc and AlSc possess the highest (\textbar{}ICOHP\textbar{} = 2.20 eV) and the lowest (\textbar{}ICOHP\textbar{} = 1.05 eV) strength, respectively. This variation in bond strength is ascribed to the differences in bond lengths, with Al$_3$Sc having the shortest (2.68 Å) and AlSc having the longest bond length (3.37 Å), respectively. The second strongest bonds in Al$_3$Sc, Al$_2$Sc, and AlSc are all the Al-Sc bonds. The \textit{p}-\textit{d} covalent hybridization between Al and Sc contributes mostly to the strength of the Al-Sc bond (Fig. \ref{fig:cohp}a$_{2}$-c$_{2}$). In terms of the Al-Sc bond, the highest strength appears in AlSc (\textbar{}ICOHP\textbar{} = 1.04 eV), while the lowest one appears in Al$_2$Sc (\textbar{}ICOHP\textbar{} = 0.62 eV), which is exactly opposite to the sequence sorted by the Al-Al bond strength. The bond length is also a crucial factor in deciding the Al-Sc bond strength. Specifically, AlSc and Al$_2$Sc with the strongest and the weakest Al-Sc bond possess the shortest (2.92 Å) and the longest (3.14 Å) length, respectively. Apart from Al-Al and Al-Sc bonds, even though the lengths of Sc-Sc bond are also less than 3.5 Å in Al$_2$Sc (3.28 Å, Fig. \ref{fig:cohp}c$_{2}$) and AlSc (3.37 Å, Fig. \ref{fig:cohp}c$_{3}$), their bonding strength is already very weak. The \textbar{}ICOHP\textbar{} values of the Sc-Sc bond in Al$_2$Sc and AlSc are calculated to be 0.32 eV and 0.28 eV, respectively, which are significantly smaller than those of their corresponding Al-Al bond. 

\begin{table*}[!ht]
\caption{\textbf{Chemical bonds with a length of less than 3.5 Å, and their respective bond length, bond number in the unit cell and \textbar{}ICOHP\textbar{}, and total \textbar{}ICOHP\textbar{} per cell and averaged \textbar{}ICOHP\textbar{} per atom of Al$_3$Sc, Al$_2$Sc and AlSc.}}
\label{tab:ICOHP}
\begin{tabular}{cccccccc}
\hline
Sys.                 & Neig.             & \begin{tabular}[c]{@{}c@{}}Bond\\ type\end{tabular} & \begin{tabular}[c]{@{}c@{}}Bond\\ length\\ (Å)\end{tabular} & \begin{tabular}[c]{@{}c@{}}\textbar{}ICOHP\textbar{}\\ (eV)\end{tabular} & \begin{tabular}[c]{@{}c@{}}Bond\\ num.\\  per\\ cell\end{tabular} & \begin{tabular}[c]{@{}c@{}}\textbar{}ICOHP\textbar{}\\ per\\ cell\\ (eV/cell)\end{tabular} & \begin{tabular}[c]{@{}c@{}}\textbar{}ICOHP\textbar{}\\ per\\ atom\\ (eV/atom)\end{tabular} \\ \hline
\multirow{2}{*}{Al$_3$Sc} & \multirow{2}{*}{1\textsuperscript{st}} & Al-Al                                               & \multirow{2}{*}{2.90}                                       & 1.63                                                   & 12                                                                  & \multirow{2}{*}{30.96}                                                   & \multirow{2}{*}{7.74}                                                    \\
                       &                      & Al-Sc                                               &                                                             & 0.95                                                   & 12                                                                  &                                                                          &                                                                          \\
\multirow{3}{*}{Al$_2$Sc} & 1\textsuperscript{st}                  & Al-Al                                               & 2.68                                                        & 2.20                                                   & 48                                                                  & \multirow{3}{*}{170.72}                                                  & \multirow{3}{*}{7.11}                                                    \\
                       & 2\textsuperscript{nd}                  & Al-Sc                                               & 3.14                                                        & 0.62                                                   & 96                                                                  &                                                                          &                                                                          \\
                       & 3\textsuperscript{rd}                  & Sc-Sc                                               & 3.28                                                        & 0.32                                                   & 16                                                                  &                                                                          &                                                                          \\
\multirow{3}{*}{AlSc}  & 1\textsuperscript{st}                  & Al-Sc                                               & 2.92                                                        & 1.04                                                   & 8                                                                   & \multirow{3}{*}{12.36}                                                   & \multirow{3}{*}{6.18}                                                    \\
                       & \multirow{2}{*}{2\textsuperscript{nd}} & Al-Al                                               & \multirow{2}{*}{3.37}                                       & 1.05                                                   & 3                                                                   &                                                                          &                                                                          \\
                       &                      & Sc-Sc                                               &                                                             & 0.28                                                   & 3                                                                   &                                                                          &                                                                          \\ \hline
\end{tabular}
\end{table*}

\section{Discussion}\label{sec:discussion}
We now discuss the underlying mechanisms behind different elastic stiffness and ductility-brittleness of Al$_3$Sc, Al$_2$Sc, and AlSc from electronic structure. In this work, the mean bonding strengths (MBS) of the different compounds are calculated. This evaluation involves two steps. Firstly, the total \textbar{}ICOHP\textbar{} per unit cell is computed by multiplying the \textbar{}ICOHP\textbar{} values of different bonds by their respective bond numbers in the unit cell and summing them up. Secondly, the averaged \textbar{}ICOHP\textbar{} per atom, which can be used to characterize the MBS, is calculated by dividing the total \textbar{}ICOHP\textbar{} by the number of atoms in the unit cell. The results are presented in Table \ref{tab:ICOHP}. We find that the averaged \textbar{}ICOHP\textbar{} of AlSc is 6.18 eV/atom, which is significantly lower than those of Al$_3$Sc (7.74 eV/atom) and Al$_2$Sc (7.11 eV/atom). This indicates that AlSc has the smallest MBS, even though the \textbar{}ICOHP\textbar{} value of the second strongest Al-Sc bond in this compound is the largest. The weakest MBS of AlSc accounts well for its lowest elastic stiffness, including \textit{B}, \textit{G} and \textit{E}, as displayed in Fig. \ref{fig:ElasticModuli_0 K}. Analyses show that the smallest MBS of AlSc arises from its weakest Al-Al bond. The \textbar{}ICOHP\textbar{} value of the Al-Al bond in AlSc is 1.05 eV, which is much smaller than those in Al$_3$Sc (1.63 eV) and Al$_2$Sc (2.20 eV). Moreover, we find that the longer bond length of the Al-Al bond in AlSc should be responsible for the weaker bond strength. Different from the first neighboring Al-Al bonds in the L1$_{2}$-type Al$_3$Sc and the C15-type Al$_2$Sc, the Al-Al bond in the B2-type AlSc is the second neighboring, leading to a much longer bond length in AlSc (3.37 Å) compared to Al$_3$Sc (2.90 Å) and Al$_2$Sc (2.68 Å) (Table \ref{tab:ICOHP}). 

It is well-known that there exists an inevitable trade-off between intrinsic strength/stiffness and ductility in materials. Materials with high strength/stiffness often exhibit poor toughness and \textit{vice versa}. For example, pure metals (\textit{e.g.}, Al and Cu) are ductile but relatively soft, whereas ceramics are hard but brittle. In this study, AlSc with the lowest stiffness possesses superior ductility, aligning well with the stiffness-ductility trade-off principle. Thus, the weakest Al-Al bond of AlSc would also contribute significantly to its exceptional intrinsic ductility. In addition, from the analyses of ELF distribution along the Al-Al bond (Fig. \ref{fig:elf2d}a), it is evident that the metallicity of the Al-Al bond in AlSc is more pronounced compared to those in Al$_3$Sc and Al$_2$Sc. This enhanced metallicity would also play a non-negligible role in the superior ductility of AlSc relative to Al$_2$Sc and Al$_3$Sc. The notable metallicity of the Al-Al bond in AlSc could be attributed to the higher valence electron concentration around Al atoms. As shown in Table \ref{tab:bader}, the Bader charge of Al in AlSc is 4.03, which is significantly higher than that of Al$_2$Sc (3.56) and Al$_3$Sc (3.34). Increasing valence electron concentration generally tends to improve the intrinsic ductility of materials \cite{RN607}. Before closing the discussion, we would like to stress that apart from the emphasized intrinsic ductility-brittleness, the plasticity of metallic materials is also influenced by defects \cite{RN625}, such as dislocation. The AlSc compound, with its simple B2 structure, may facilitate dislocation movement. Details of dislocation structures of Al$_3$Sc, Al$_2$Sc and AlSc, out of the scope of this work, need to be further investigated specifically.

\begin{table*}[!ht]
\caption{\textbf{The absolute values of the integration of the orbital-resolved COHP (\textbar{}ICOHP\textbar{}) of Al$_3$Sc, Al$_2$Sc and AlSc.}}
\label{tab:pCOHP}
\begin{tabular}{cccccccccc}
\hline
\multicolumn{2}{c}{\multirow{2}{*}{Type}}                                              & \multicolumn{2}{c}{Al$_3$Sc} & \multicolumn{3}{c}{Al$_2$Sc} & \multicolumn{3}{c}{AlSc} \\ \cline{3-10} 
\multicolumn{2}{c}{}                                                                   & Al-Al       & Al-Sc       & Al-Al   & Al-Sc  & Sc-Sc  & Al-Al  & Al-Sc  & Sc-Sc  \\ \hline
\multirow{7}{*}{\begin{tabular}[c]{@{}c@{}}\textbar{}ICOHP\textbar{}\\ (eV)\end{tabular}} & \textit{s-s} & 0.02        & 0.07        & 0.03    & 0.04   & 0.02   & 0      & 0.16   & 0.03   \\
                                                                        & \textit{s-p} & 0.77        & 0.16        & 1.12    & 0.14   & —      & 0.4    & 0.21   & —      \\
                                                                        & \textit{p-p} & 0.84        & —           & 1.05    & —      & —      & 0.65   & —      & —      \\
                                                                        & \textit{s-d} & —           & 0.2         & —       & 0.11   & 0.12   & —      & 0.21   & 0.09   \\
                                                                        & \textit{p-d} & —           & 0.5         & —       & 0.33   & —      & —      & 0.46   & —      \\
                                                                        & \textit{d-d} & —           & —           & —       & —      & 0.18   & —      & —      & 0.17   \\
                                                                        & Total        & 1.63        & 0.95        & 2.2     & 0.62   & 0.32   & 1.05   & 1.04   & 0.28   \\ \cline{1-10}
\end{tabular}
\end{table*}

\section{Conclusions}
In this work, the stable phases, the ground-state and finite-temperature inherent ductility-brittleness and the electronic structures of the Al-Sc binary systems with Sc content of less than 50 at. \% are studied systematically. Using the variable-composition evolutionary structure search algorithm, it is confirmed that apart from the L1$_{2}$-type Al$_3$Sc, the C15-type Al$_2$Sc and the B2-type AlSc, there are no other stable Al-Sc binary intermetallics. From all Pugh’s, Pettifor’s, and Poisson’s ductility-brittleness criteria, Al$_3$Sc and Al$_2$Sc are inherently brittle at ground state. AlSc possesses a prominently superior intrinsic ductility compared to Al$_3$Sc and Al$_2$Sc evaluated from all Pugh’s, Pettifor’s and Poisson’s criteria. Through AIMD simulation, the finite-temperature elastic moduli are determined. By rising temperature, for all Al$_3$Sc, Al$_2$Sc, and AlSc, their intrinsic ductilities can be notably improved. As the temperature increases, accompanied by the transition of the Cauchy pressure from negative to positive, AlSc is unequivocally classified as the inherently ductile based on all the criteria considered. However, Al$_3$Sc and Al$_2$Sc retain their brittle nature even at temperatures up to 1200 K. In all Al$_3$Sc, Al$_2$Sc and AlSc, the Al-Al bond, resulting from \textit{s}-\textit{p} and \textit{p}-\textit{p} orbital hybridizations, and the Al-Sc bond, dominated by \textit{p}-\textit{d} covalent hybridization, are the first and the second strongest chemical bonds, respectively. The mean bond strength (MBS) is introduced in this work to explain the differences in intrinsic mechanical properties of Al$_3$Sc, Al$_2$Sc, and AlSc. The weaker Al-Al bond in AlSc, leading to a smaller MBS, could be the origin of the softer elastic stiffness and superior intrinsic ductility. The longer length of the Al-Al bond in AlSc is responsible for its weaker bond strength. Moreover, an enhanced metallicity of the Al-Al bond in AlSc would also play a non-negligible role in its superior ductility. The findings of this work are expected to provide a crucial foundation for the design of alloy compositions and optimization of thermomechanical processing for Al-Sc target materials.

\appendix
\section{Appendix}\label{appen:1}
\setcounter{equation}{0}
\renewcommand{\theequation}{A\arabic{equation}}
The isotropic elastic moduli of bulk modulus \textit{B}, shear modulus \textit{G}, Young’s modulus \textit{E} and Poisson’s ratio \textit{$\nu$} are calculated by Voigt-Reuss-Hill (V-R-H) approximation \cite{RN396} as follows:
\begin{equation}
B={\frac{1}{2}}(B_{\mathrm{V}}\!+\!B_{\mathrm{R}})
\end{equation}
\begin{equation}
G={\frac{1}{2}}\left(G_{\mathrm{V}}+G_{\mathrm{R}}\right)
\end{equation}
\begin{equation}
V\!=\!{\frac{3B\!-\!2G}{2(3B\!+\!G)}}
\end{equation}
\begin{equation}
E={\frac{9G B}{G+3B}}
\end{equation}
\\ %
where \textit{B}$_V$ and\textit{G}$_V$ are Voigt bulk modulus and Voigt shear modulus, respectively, and \textit{B}$_R$ and \textit{G}$_R$ are Reuss bulk modulus and Reuss shear modulus, respectively. For cubic crystals, \textit{B}$_V$,\textit{G}$_V$, \textit{B}$_R$ and\textit{G}$_R$ can be calculated by the relations: 
\begin{equation}
G_{\mathrm{V}}\!=\!{\frac{1}{5}}\left[(C_{11}\!-\!C_{12})\!+\!3C_{44}\right]    
\end{equation}
\begin{equation}
B_{\mathrm{V}}=\frac{1}{3}\left(C_{11}{+}2C_{12}\right)    
\end{equation}
\begin{equation}
G_{\mathrm{R}}\!=\frac{\,5C_{44}(C_{11}\mathrm{-}C_{12})\,}{3(C_{11}\mathrm{-}C_{12})+4C_{44}}\,   
\end{equation}
\begin{equation}
B_{\mathrm{R}}\!=\!\frac{1}{3}\left(C_{11}\!+\!2C_{12}\right)   
\end{equation}

\section{Acknowledgments}
This work is supported by the National Key R\&D Program of China (2022YFB3504401).

\bibliographystyle{apsrev4-1}
\bibliography{ref}
\end{document}


\title{\textbf{Supplementary Materials to} \\ "Finite-temperature ductility-brittleness and electronic structures of Al$_{n}$Sc (n=1, 2 and 3)"}

\author{Xue-Qian Wang}
\affiliation{%
Key Laboratory for Anisotropy and Texture of Materials (Ministry of Education), School of Material Science and Engineering, Northeastern University, Shenyang 110819, China.
}%
\author{Ying Zhao}
\affiliation{%
Key Laboratory for Anisotropy and Texture of Materials (Ministry of Education), School of Material Science and Engineering, Northeastern University, Shenyang 110819, China.
}%
\author{Hao-Xuan Liu}
\affiliation{%
Key Laboratory for Anisotropy and Texture of Materials (Ministry of Education), School of Material Science and Engineering, Northeastern University, Shenyang 110819, China.
}%
\author{Shuchen Sun}
\email{sunsc@smm.neu.edu.cn}
\affiliation{%
School of Metallurgy, Northeastern University, Shenyang 110819, China.
}%
\author{Hongbo Yang}
\email{yanghongbo203@grirem.com}
\affiliation{%
Rare Earth Functional Materials (Xiong'an) Innovation Center Co., Ltd., Xiong'an 071700, China
}%
\author{Jiamin Zhong}
\affiliation{%
Rare Earth Functional Materials (Xiong'an) Innovation Center Co., Ltd., Xiong'an 071700, China
}%
\author{Ganfeng Tu}
\affiliation{%
School of Metallurgy, Northeastern University, Shenyang 110819, China.
}%
\author{Song Li}
\affiliation{%
Key Laboratory for Anisotropy and Texture of Materials (Ministry of Education), School of Material Science and Engineering, Northeastern University, Shenyang 110819, China.
}%
\author{Hai-Le Yan}
\email{yanhaile@mail.neu.edu.cn}
\affiliation{%
Key Laboratory for Anisotropy and Texture of Materials (Ministry of Education), School of Material Science and Engineering, Northeastern University, Shenyang 110819, China.
}%
\author{Liang Zuo}
\affiliation{%
Key Laboratory for Anisotropy and Texture of Materials (Ministry of Education), School of Material Science and Engineering, Northeastern University, Shenyang 110819, China.
}%

	\maketitle

	\begin{figure}
		\begin{center}
			\includegraphics[width=0.80\textwidth, clip]{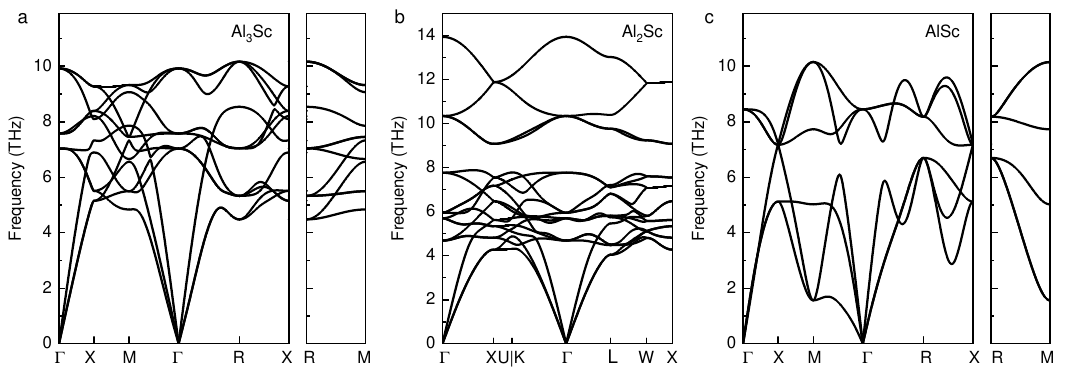}
		\end{center}
		\caption{\textbf{Phonon dispersion curves of Al$_3$Sc (a), Al$_2$Sc (b) and AlSc (c) with the ground-state structures.}}
		\label{Fig.S1}
	\end{figure}

	\begin{figure}
		\begin{center}
			\includegraphics[width=0.80\textwidth, clip]{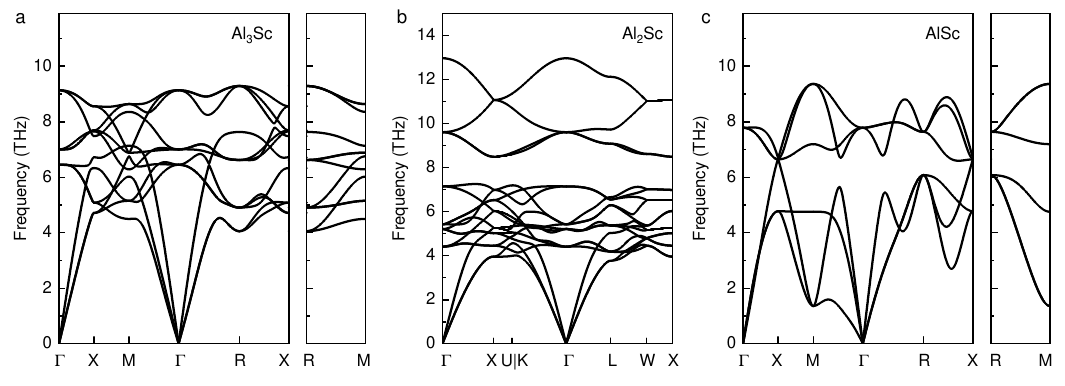}
		\end{center}
		\caption{\textbf{Phonon dispersion curves of Al$_3$Sc (a), Al$_2$Sc (b) and AlSc (c) with the equilibrium structures at 1200 K.}
		}
		\label{Fig.S2}
	\end{figure}